%% file: main.tex
\documentclass[sigconf,nonacm,dvipsnames]{acmart}

\usepackage[utf8]{inputenc}
\input{meta/packages}
\input{meta/macros}

\title{Payout Races and Congested Channels: A Formal Analysis of Security in the Lightning Network}
\input{meta/authors}

\begin{document}

\input{sections/abstract}
\maketitle

\input{sections/intro}
\input{sections/background}
\input{sections/model}
\input{sections/props}
\input{sections/results}
\input{sections/related}
\input{sections/conclusion}

\input{sections/acks}

\bibliographystyle{plainnat}
\bibliography{main}

\appendix
\input{sections/formal-methods}
\input{sections/model-details}

\end{document}

%% file: meta/packages.tex
\microtypecontext{spacing=nonfrench}

\usepackage{adjustbox}
\usepackage{booktabs}
\usepackage{circledsteps}
\usepackage{paralist}
\usepackage[group-digits=integer,group-minimum-digits=4,group-separator={{,}},per-mode=symbol,detect-all]{siunitx}
\usepackage{xspace}
\usepackage{tcolorbox}

%% file: meta/macros.tex
\hypersetup{
     colorlinks=true,
     citecolor=blue,
     linkcolor=black,
}

\newif\ifDraft

\Draftfalse

\usepackage[lambda,adversary,asymptotics,sets,landau,probability,operators]{cryptocode}

\newcommand{\addText}[1]{%
  \ifDraft\textcolor{blue}{#1}\else#1\fi
}

\makeatletter
\g@addto@macro{\UrlBreaks}{\UrlOrds}
\makeatother

\usepackage[lambda,adversary,asymptotics,sets,landau,probability,operators]{cryptocode}
\usepackage{cleveref}

\newtheorem{property}{Property}

\bgroup\fboxsep=0pt

\NewEnviron{dottedbox}{
\par
\begin{tikzpicture}
\node[rectangle,minimum width=0.9\textwidth] (m) {\begin{minipage}{\textwidth}\BODY\end{minipage}};
\draw[dashed] (m.south west) rectangle (m.north east);
\end{tikzpicture}
}

\definecolor{gamechangecolor}{gray}{0.90}

\tcbuselibrary{theorems}
\tcbuselibrary{skins}
\makeatletter
\crefformat{tcb@cnt@protocol}{protocol~#2#1#3}
\Crefformat{tcb@cnt@Protocol}{Protocol~#2#1#3}

\makeatother
\newtcbtheorem[auto counter]{protocol}{Protocol}{
	left=8pt,
	right=8pt,
	top=0pt,
	bottom=0pt,
    fonttitle=\normalsize,
    fontupper=\footnotesize,
	coltitle=black!,
	colbacktitle=black!0,
	colback=black!0, 
	boxrule=0.5pt,
	colframe=black!100,
	coltext=black!100,
	arc=0pt
}{proto}

\makeatother
\newtcbtheorem[auto counter]{functionality}{Functionality}{
	left=8pt,
	right=8pt,
	top=0pt,
	bottom=0pt,
    fonttitle=\bfseries\normalsize,
    fontupper=\small,
	coltitle=black!,
	colbacktitle=black!0,
	colback=black!0, 
	boxrule=0.5pt,
	colframe=black!100,
	coltext=black!100,
	arc=0pt
}{func}

\crefformat{tcb@cnt@functionality}{functionality~#2#1#3}
\Crefformat{tcb@cnt@Functionality}{Functionality~#2#1#3}

\usepackage{seqsplit}
\newcommand{\ttt}[1]{%
  \begingroup
    \protect\renewcommand{\seqinsert}{\ifmmode\allowbreak\else\-\fi}%
    \protect\texttt{\protect\seqinsert{\protect\seqsplit{\small#1}}}%
  \endgroup
}
\newcommand{\tttscript}[1]{%
  \begingroup
    \protect\renewcommand{\seqinsert}{\ifmmode\allowbreak\else\-\fi}%
    \protect\texttt{\protect\seqinsert{\protect\seqsplit{\scriptsize#1}}}%
  \endgroup
}
\newcommand{\tttfoot}[1]{%
  \begingroup
    \protect\renewcommand{\seqinsert}{\ifmmode\allowbreak\else\-\fi}%
    \protect\texttt{\protect\seqinsert{\protect\seqsplit{\footnotesize#1}}}%
  \endgroup
}

\usetikzlibrary{shapes, automata, positioning, arrows, calc}
\tikzset{
    >=stealth, % makes the arrow heads bold
    node distance=3.5cm, % specifies the minimum distance between two nodes. Change if necessary.
    every state/.style={rectangle, thick, fill=gray!10}, % sets the properties for each ’state’ node
    initial text=$ $, % sets the text that appears on the start arrow
}

\newcommand{\buchi}{B{\"u}chi\xspace}
\renewcommand{\ln}{Lightning Network\xspace}

\colorlet{myred}{red!50!white}
\newcommand{\spin}{\textsc{Spin}\xspace}
\newcommand{\promela}{\textsc{Promela}\xspace}

\newcommand{\funded}{\texttt{FUNDED}\xspace}

\newcommand{\fail}{\texttt{FAIL\_CHANNEL}\xspace}

\newcommand{\add}{\ttt{UPDATE\_ADD\_HTLC}\xspace}
\newcommand{\comm}{\ttt{COMMITMENT\_SIGNED}\xspace}
\newcommand{\revokeandack}{\ttt{REVOKE\_AND\_ACK}\xspace}
\newcommand{\fulfill}{\ttt{UPDATE\_FULFILL\_HTLC}\xspace}
\newcommand{\error}{\ttt{ERROR}\xspace}
\newcommand{\failhtlc}{\ttt{UPDATE\_FAIL\_HTLC}\xspace}
\newcommand{\malformedhtlc}{\ttt{UPDATE\_FAIL\_MALFORMED\_HTLC}\xspace}

\newcommand{\addshort}{\ttt{ADD}\xspace}
\newcommand{\commshort}{\ttt{COMM}\xspace}
\newcommand{\revokeandackshort}{\ttt{REV}\xspace}
\newcommand{\fulfillshort}{\ttt{FULF}\xspace}
\newcommand{\errorshort}{\ttt{ERR}\xspace}
\newcommand{\failhtlcshort}{\ttt{FAIL}\xspace}
\newcommand{\malformedhtlcshort}{\ttt{FAILM}\xspace}

\newcommand{\maxhtlcs}{\ttt{max\_accepted\_htlcs}\xspace}

\newcommand*{\step}[1]{\Circled[fill color=black,inner color=white]{#1}}

\newcommand{\lnd}{\texttt{lnd}\xspace}
\newcommand{\attack}{Payout Race\xspace}

%% file: meta/authors.tex
\author{Ben Weintraub}
 \affiliation{%
   \institution{Northeastern University}
   \country{}
 }

\author{Satwik Prabhu Kumble}
 \affiliation{%
   \institution{TU Delft}
   \country{}
 }
 
\author{Cristina Nita-Rotaru}
 \affiliation{%
   \institution{Northeastern University}
   \country{}
 }
 
\author{Stefanie Roos}
 \affiliation{%
   \institution{University of Kaiserslautern-Landau}
   \country{}
 }

%% file: sections/abstract.tex
\begin{abstract}
The Lightning Network, a payment channel network with a market cap of over 192M USD, is designed to resolve Bitcoin's scalability issues through fast off-chain transactions. 
There are multiple Lightning Network client implementations, all of which conform to the same textual specifications known as BOLTs. 
Several vulnerabilities have been manually discovered, but to-date there have been few works systematically analyzing the security of the Lightning Network.

In this work, we take a foundational approach to analyzing the security of the Lightning Network with the help of formal methods.
Based on the BOLTs' specifications, we build a detailed formal model of the Lightning Network's \addText{single-hop} payment protocol and verify it using the \spin model checker.
\addText{Our model captures both concurrency and error semantics of the payment protocol.}
We then define several security properties which capture the correct intermediate operation of the protocol, ensuring that the outcome is always certain to both channel peers,
and using them
we re-discover
a known attack previously reported in the literature along with a novel attack, referred to as a \attack.
A \attack consists of a particular sequence of events that can lead to an ambiguity in the protocol in which innocent users can unwittingly lose funds.
We confirm the practicality of this attack by reproducing it in a local testbed environment.
\end{abstract}

%% file: sections/intro.tex
\section{Introduction}

Scalability has long been a struggle for blockchain-based digital currencies~\cite{Rizzo_2016}.
\emph{Payment channel networks} (PCN) are one popular solution to this problem.
PCNs facilitate participants making payments off-chain, while still maintaining the security properties of the underlying blockchain.
The largest PCN in deployment is the \ln~\cite{poon2016bitcoin} with a peer count of over \num{49000} and a market cap of over 192M USD\footnote{1ml.com/statistics}. %\stef{have these numbers been updated? I see more than 200M USD on the 1ML page}
The \ln (LN) relies on Bitcoin~\cite{nakamoto2008bitcoin} as its underlying blockchain.

The \ln operates in three phases: channel establishment, channel operation, and channel closure.
In the channel establishment phase, two peers escrow funds between them in a transaction whose output they can only spend with a transaction signed by both peers.
In the channel operation phase, the two peers create a new transaction that pays the escrowed funds to the peers.
The transaction changes the ratio of how much of the funds goes to each peer and represents the updated channel balance as a result of the payment. 
The two peers each sign the new transaction with their respective private keys and each send the resulting signature to the other peer, so that both peers have their own signature for the transaction as well as the other peer's signature.
Neither peer publishes this transaction on the blockchain.
They store the transaction locally.
The transaction serves as protection for each peer.
At any time, they can submit the transaction to the blockchain network and retrieve their funds.
This funds-retrieval process is called channel closure---the third phase.
This protocol was designed by \citet{poon2016bitcoin} and later standardized in the form of a specification documents called the Basis of Lightning Technology (BOLT)~\cite{bolts}.

As a result of its growing popularity, the \ln has been subject to numerous attacks in the literature
such as the Flood \& Loot attack~\cite{harris2020flood}, griefing attacks~\cite{Weintraub_Nita-Rotaru_Roos_2021, discharged, perez2020lockdown, interledger, mizrahi2021congestion}, the Wormhole attack~\cite{wormhole}, which allows intermediaries to steal the fees of others, as well as various privacy attacks, which allow parties to infer sensitive information about user identities~\cite{kumble2021lightning,kumar2023anonymity, rohrer2020counting, nisslmueller2020active, tikhomorov_privacy_2020, kappos2021empirical} or the amount of funds users own~\cite{herrera2019difficulty,tikhomorov_probing_2021,nisslmueller2020active,kappos2021empirical}.

The \ln's use as a financial medium of exchange underscores the importance of its security, and makes finding any latent security issues a top priority. 
Most prior attempts to find vulnerabilities in the \ln have done so via manual analysis by domain experts---a difficult and time-consuming process.
Only a few works have focused on ensuring the \emph{absence} of vulnerabilities.
\citet{Rain_game_2023} created a game-theoretic model of the LN, but their model is based on the original 2016 Lightning paper by \citet{poon2016bitcoin} and does not reflect the significant changes to the protocol developed since then.
Additionally, they model only the channel closure and overlay routing, which ---while valuable --- are not the phases in which most network communication occurs.
Likewise, \citet{Kiayias_Litos_2020} built a theoretical model of the cryptography of the LN based on the BOLTs.
This, however, only works under assumptions of idealized functionality 
and thus does not reflect incorrect behavior that can result under more realistic network and operational conditions.
Meanwhile, specific implementations of the \ln such as \lnd~\cite{lnd}, \texttt{CoreLightning}~\cite{corelightning}, and \texttt{Eclair}~\cite{eclair} are subject to unit tests and code reviews to find and prevent mistakes in said implementations.

One approach for getting assurance about the correctness of systems is using formal methods.
Unlike testing, formal methods can help disambiguate system specifications and can expose flaws in system requirements, often not captured through testing. 
They also provide mathematical proofs about the system behavior.
Formal methods have been shown to be successful for uncovering security flaws in many network protocols including TCP~\cite{von2020automated, Bishop_Fairbairn_Mehnert_Norrish_Ridge_Sewell_Smith_Wansbrough_2019}, TLS~\cite{bhargavan2017verified, delignat2017implementing, cremers2017comprehensive}, Bluetooth~\cite{arai2014formal, chang2007formal, nguyen2014formal, phan2012analyzing, wu2022formal}, and 5G~\cite{basin2018formal}, and have also seen similar success in cryptographic protocols such as Signal~\cite{kobeissi2017automated, cremers2020clone}, Noise~\cite{girol2020spectral, kobeissi2019noise}, Needham-Schroeder~\cite{maggi2002using, lowe1995attack, lowe1996breaking}, and Diffie-Hellman~\cite{cremers2019prime}.

The use of formal methods in the blockchain space has been limited, however.
A number of works have explored the correctness of smart contracts---in the same vein as program analysis~\cite{Amani_verifying_2018, Annekov_concert_2020, Sun_formal_2020, grishchenko_ethertrust_2018, Bai_formal_2018, Osterland_model_2020, Almakhour_formal_2023, Nehai_model_2018, Mavridou_verisolid_2019, Nam_formal_2022},
one paper explored the correctness of one of Bitcoin's core protocols~\cite{Modesti_Shahandashti_McCorry_Hao_2021}, and another modeled byzantine fault tolerance in the Red Belly blockchain~\cite{crain2021red,tholoniat2022formal}.
However, none of these works, nor any others to our knowledge, have sufficiently addressed concerns about the correctness of the LN and its associated protocols.

To our knowledge, only one work has applied formal methods to the \ln~\cite{Grundmann_Hartenstein_2023}.
However, this work abstracts away several important details, which may be the reason why it does not find any vulnerabilities. 
First, they only model a single payment at a time, which overlooks the complexities of concurrency.
Second, they do not model any error conditions, which are important because they represent ways in which the protocol could exit unexpectedly without executing necessary clean up actions.
Finally, they define only a single security property, which is too abstract to model any unexpected behavior within the protocol---even if no funds are lost.
Additionally, the work is not peer reviewed and the only public artifacts are a short two-page paper~\cite{Grundmann_2021} \addText{and an extended progress report~\cite{Grundmann_Hartenstein_2023}},
thus hindering the development of a cumulative literature.

In this work, we build a formal model of \addText{single-hop transactions in} the \ln payment protocol.
We model this protocol as a \emph{finite state machine} (FSM), as well as provide five security-critical properties that are necessary to protect users' funds and their accessibility thereof.
These properties capture not only the overall successful completion of payments, but also numerous error states not modeled in previous work.
We also define properties that capture the correct intermediate operation of the protocol, thus ensuring that there are no states in which a peer's view of the transaction outcome is uncertain.
The model and the properties were informed by the official \ln BOLTs.
We then formally verified this model using the \spin model checker to explore all protocol executions and confirm if any such executions violate our security properties.
We find that two of the five properties can be violated. The first violation is a reproduction of the previously published \textit{congestion attack}~\cite{mizrahi2021congestion}.
The second violation lead us to discover a novel attack against the system in which an in-flight payment can end up in an ambiguous state allowing either party to claim the funds---possibly against the expectations of the other peer on the channel (i.e., the counterparty).
We call this a \attack attack, and we reproduced this attack in a local testbed environment.
We discuss mitigations and impact of the attack, as well as limitations of the mitigations techniques due to message timing in distributed systems.

\paragraph{Ethics}
The vulnerability of a financial network is a sensitive matter.
While developing stronger defenses to this attack is out of the scope of this work, we have alerted the LN's security contact point of our findings and the risks.
No perfect defense is possible at this time without a massive redesign of the protocol.
However, the LN developers are considering clarifications to the specification documents to inform protocol implementers of various risk factors to be aware of.
One popular LN client, \lnd, has implemented features in an attempt to limit exposure to this attack.
While these mitigations are not a panacea, they do increase the attack difficulty by narrowing the vulnerable time window and might make good candidates for standardization.
Though they are all limited by both lack of developer awareness as well as unrealistic uptime expectations.
We highlight these efforts and their drawbacks in detail in \Cref{sec:results}.

\paragraph{Contributions}
Our contributions are the following
\begin{itemize}
  \item We formally model \addText{single-hop transactions in} the \ln payment protocol as a finite state machine as well as define five security properties. We open source our model and properties for the benefit of the \ln and security communities\footnote{\url{https://zenodo.org/records/11002329}}.
  \item We show that two of the properties can be violated, thus reproducing a prior attack and leading us to discover the new \attack attack.
  \item We show that our novel \attack attack is a credible threat by reproducing it in a testbed environment consisting of \lnd clients. We confirm that \lnd is following the BOLT specifications via manual code inspection. 
\end{itemize}

%% file: sections/background.tex
\section{Background}\label{sec:background}

\emph{Payment channel networks} (PCN) are a type of financial payment system designed to increase throughput and privacy of blockchain transactions---both of which are salient issues among digital currencies, including Bitcoin.
The \ln (LN) is a PCN implementation that uses Bitcoin as its underlying blockchain~\cite{poon2016bitcoin}.

\subsection{The UTXO Model of Transactions}
As the \ln relies on the operation of Bitcoin, it is important to understand the mechanics of Bitcoin transactions.
Bitcoin follows the \emph{unspent transaction output} (UTXO) model for storing and transferring its eponymous digital currency.
In the UTXO model, tokens are not fungible.
Technically speaking, a user does not spend bitcoins (BTC) directly; a user spends the output of a previously mined transaction---however many BTC that entails.
A transaction can have multiple inputs and multiple outputs.
We define an output as $\theta := (amount,\phi)$, where $amount$ is the value (in BTC for LN), and $\phi$ is some condition that must be satisfied to spend that output.
For the case of Bitcoin, we define the condition $\phi$ as a function that accepts some input, and must evaluate to $true$.
One important condition for our work is \textit{transaction maturity}, which is defined as the Bitcoin block number of a future block.
It can be specified in relative or absolute terms, and it demarcates a point in time before which a transaction is invalid. 
A transaction will not be included in a block by an honest miner if it is not yet mature.

We define an input of a transaction as $\tau := (ref(\theta), args)$.
That is, an input includes a reference to an output along with a vector $args$ of input arguments that must satisfy the output condition $\theta.\phi$.

\subsection{Payment Channels}
Payment channel networks (PCNs) consist of a network of interlinked \emph{payment channels}.
A payment channel is a blockchain-backed financial agreement between two or more parties, which allows them to make verifiable payments to each other without requiring that all of the transactions are published on the blockchain.
In the \ln, a channel must have exactly two parties.
The party connected to a peer via a payment channel is called the \emph{counterparty}.
This terminology is symmetric and simply indicates whose perspective we are taking at the moment.

\paragraph{Channel Opening and Closing}
A payment channel in LN is established by one of the channel parties escrowing funds in a blockchain transaction that contains an output condition $\phi_{close}$ requiring 2-of-2 signatures to release the funds back to their rightful owner.
This is called a \emph{funding transaction}.
As part of the channel establishment process, both parties cooperate to form a signed blockchain transaction, which they broadcast to the Bitcoin network.
The channel is considered \emph{funded} when the funding transaction has been added to the ledger and had some configurable number of blocks mined on top of it.

The two parties then cooperate to form a \emph{closing transaction}.
This is a transaction that includes an input $\tau_{close}$ that contains both parties' signatures in the arguments; these are both needed to satisfy $\phi_{close}$.
Creating this closing transaction is done as part of channel establishment,
but they do not immediately send it to the blockchain network for processing.
Instead, both parties store the transaction locally.
At any time, either party can submit the closing transaction to the blockchain to release the escrowed funds.
The closing transaction can also be updated as discussed next.

\paragraph{Channel Operation}\label{sec:bkg:chan_op}
When the local party wants to make a payment to the counterparty, they send the counterparty a \emph{commitment}---a signed blockchain transaction that spends the funding transaction output and pays out the channel balance in a potentially different ratio than was initially agreed upon.
In the \ln, this message is called \comm (\Cref{tab:msgs}, row 2).
An astute reader will notice that the closing transaction from the previous paragraph also spent the funding transaction output.
That is because this commitment---and any that follow---are replacements for the initial closing transaction.
The ability to spend from the funding transaction outputs is precisely what makes it a commitment, because the counterparty can close the channel with said commitment and receive the dictated amount.
Both parties can continue sending each other updated commitments for new channel balances---arbitrarily many times---until the channel is closed.
Each of the commitments transmitted between the peers is a valid blockchain transaction, which means that as the parties make payments to each other, they are collecting a series of valid transactions.

\paragraph{Secure Payments}
The protocol necessitates that old commitments be rendered unspendable, but enforcing this is non-trivial.
There are a number of proposed solutions to this problem~\cite{layer2sok}.
The solution used in LN is called \emph{replace-by-revocation} (RbR).
In RbR, every commitment includes a conditional penalty payout that can be triggered by providing a commitment-specific \emph{revocation key}.
After receiving a new commitment, the local party sends the counterparty a revocation key for the previous commitment.
In the \ln, this message is called \revokeandack (\Cref{tab:msgs}, row 3).
If either peer submits a transaction to the blockchain that they had previously revoked, then the counterparty has recourse by submitting a new penalty transaction that includes the acquired revocation key.
This revocation functionality is built into the $\phi$ conditions of each output in the commitment.
The penalty transaction allows the penalizing party to spend both the payout that would normally have gone to them plus the payout that would have gone to the counterparty, thus allowing them to pay the full amount of the channel to themselves.
This mechanism is supposed to disincentivize any attempts at duplicity from either peer.

\begin{figure}[h]
     \centering
     \includegraphics[width=\linewidth]{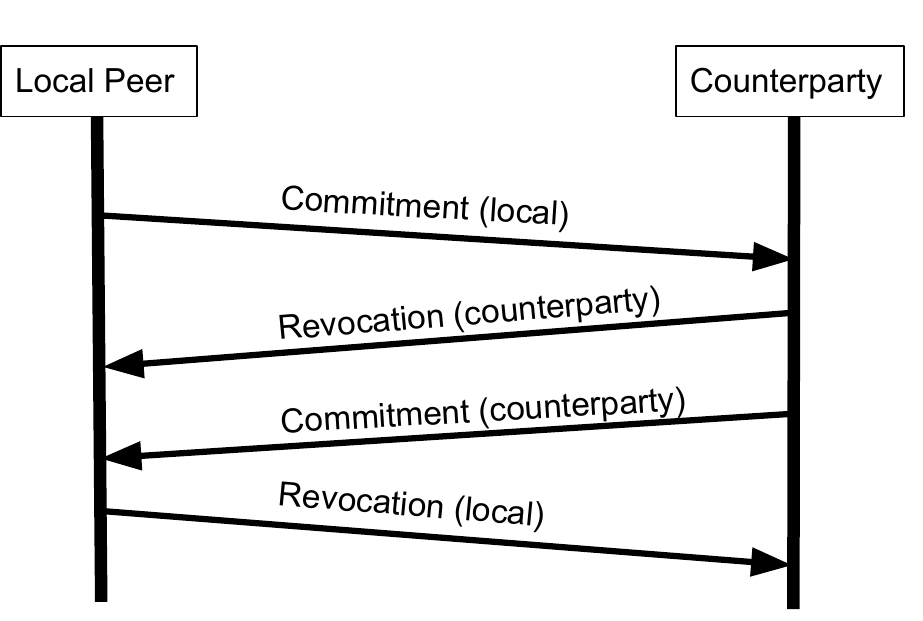}
     \caption{Each Lightning peer sends a commitment and revocations to its counterparty.}
     \label{fig:commflow}
 \end{figure}

A closing transaction must include a signature from both parties, thus each commitment message includes the signature of the sender.
This means a party does not update its view of the channel balance until they receive a new commitment signature from the counterparty.
In practice, this means that both parties must send a commitment of the new channel balance and both parties must revoke the previous balance (\Cref{fig:commflow}).

To prevent a race between the two parties to spend the outputs of a blockchain-published commitment, a timeout measured in blocks is enforced as part of the output conditions $\phi$ for the transaction submitter.
So while the non-submitting party
with a revocation key can spend the commitment output immediately, the transaction-submitting party must wait $T$ blocks before they can spend the commitment transaction outputs.
This delay is designed to give the non-transaction submitter time to publish a penalty transaction to the blockchain.

\subsection{Payment Channel Networks}
A natural limitation of the payment channel construct is that it requires a party to escrow funds for each channel it establishes.
Thus the number of diverse parties whom one can pay is limited by the liquidity of each party.
A payment channel network addresses this limitation by allowing individual payments to traverse multiple, sequentially linked payment channels.
The payment channels thus form a peer-to-peer network where any peer in the network can pay any other peer as long as there is a path in the network graph between them \emph{and} each channel on that path has enough liquidity to support the payment in question.

\paragraph{Hashed Timelock Contracts (HTLC)} A single payment that traverses multiple payment channels consists of independent transactions on each channel.
These component transactions must remain collectively atomic---either all of them complete successfully, or none of them do.
One mechanism for ensuring this atomicity is the HTLC~\cite{poon2016bitcoin}.

An HTLC payment is a three-step process.
In the first step, the payee $U_{n}$
sends the payer $U_{0}$ an invoice that includes the hash $H(p)$ of a random number $p$ and a payment amount $v$.
In the next step, each peer starting with $U_0$, sends a new HTLC request (called \add, \Cref{tab:msgs}, row 1) and a commitment for $v$ to the next peer along the payment path.
The commitments,
which are Bitcoin transactions signed by each sender,
each contain an outputs with condition $\phi$ that can only be satisfied if the submitter can also provide a preimage $p$ that hashes to $H(p)$.
In the final phase,
the payment recipient $U_{n}$ releases the preimage $p$ to its counterparty $U_{n-1}$ along the final channel in the payment.
Since the final hop's transaction is complete, peer $U_{n-1}$ can pass the preimage to peer $U_{n-2}$, who passes to $U_{n-3}$, etc., until the initial payer $U_{0}$ has received the preimage.

At this point, the payment is considered functionally complete as all parties along the payment path can redeem their updated channel funds by submitting their latest commitment to the blockchain.
The \ln protocol does, in fact, dictate an additional exchange of commitments and revocation on each channel.
This exchange exists for two reasons. First, it reduces the complexity of the redeem scripts $\phi$, thus reducing the transaction size---an important concern for the ever-growing Bitcoin blockchain~\cite{Rizzo_2016}.
Second, it provides a quicker spending path for unilaterally closed channels, which becomes possible because both parties will have assurance that the other party has received the commitment for the new amount. See the third and fourth messages in \Cref{fig:commflow}.

\begin{table*}[ht]
  \centering
  \begin{adjustbox}{width=\linewidth,center}
  \begin{tabular}{ l l p{0.5\linewidth} }
  \toprule
  \textbf{Message Type} & \textbf{Short Name} & \textbf{Purpose}\\
  \midrule
  \add & \addshort & Sent by a peer to start a new payment.\\
  \hline
  \comm & \commshort & The signatures needed for the receiving peer to have a redeemable Bitcoin transaction with the updated channel balance.\\
  \hline
  \revokeandack & \revokeandackshort & Shares the revocation key for the previous commitment. This allows the counterparty to redeem the penalty payout if the local party publishes an old commitment.\\
  \hline
  \fulfill & \fulfillshort & Includes the hash preimage that satisfies the HTLC output script $\theta.\phi$. \\
  \hline
  \error & \errorshort & An error message that can be sent by either peer at any time.\\
  \hline
  \failhtlc & \failhtlcshort & An error message to send when a \add message contains illegal or incorrect parameters.\\
  \hline
  \malformedhtlc & \malformedhtlcshort & An error message to send when a \add message cannot be parsed.\\
  \bottomrule
  \end{tabular}
  \end{adjustbox}
  \caption{The messages that LN peers can send during channel operation.}
  \label{tab:msgs}
\end{table*}

\subsection{Our Approach}
In this paper, we create a formal model of the popular PCN, the \ln.
Specifically, we model Lightning's payment protocol as defined in the public specification (i.e. Basis of Lightning Technology, BOLTs \cite{bolts}).

We opt for model checking because it guarantees a search through all possible executions, and its ability to report violating traces.
In contrast, fuzzing is useful for efficiently exploring large state spaces, but it notably does not offer any guarantees of finding violations if they exist.
Additionally, while fuzzing can find vulnerabilities in implementations, it is not well-suited to English-language specifications.

Theorem provers such as \textsc{ACL2}~\cite{acl2}, \textsc{Coq}~\cite{coq}, and \textsc{Alloy}~\cite{alloy} are also not suitable for our use case.
Like fuzzers, theorem provers require an implementation.
This implementation must be written in the language of the prover---a significant effort for the modeler.
Furthermore, while a theorem prover can verify similar properties as a model checker, and in some cases more properties, this comes at a cost.
The ability to model an interactive Turing machine means that some properties cannot be verified without human guidance.
On the other hand, model checking can be fully automated.

Model checkers are designed for exploring transition systems such as finite state machines.
We opt to use the \spin model checker, a state-of-the-art model checker designed for verifying concurrent programs~\cite{spin}. 
\addText{
Model checkers like \textsc{Tamarin}~\cite{tamarin} and \textsc{ProVerif}\cite{proverif} are designed to check cryptographic semantics and security protocols. For example,  they can model cryptographic games that capture definitions of security such as IND-CPA.
The attacks that we found in this work (\Cref{sec:results}) cannot be defended against with cryptography, and thus would not have been found by checking the cryptographic semantics.
\spin, on the other hand, is useful for modeling network protocol semantics. 
As this work focuses on network protocol semantics, we selected \spin, and ultimately, this allows us to model the network communication elements of the LN protocol without the cost of verifying the cryptographic elements. 
}
\spin also supports modelling \buchi automata, which are infinite executions of protocols.
This is critical for our model since there is no limit to the number of payments the parties can make on a single payment channel, which naturally leads to the existence of infinite executions.
For writing properties, \spin supports verification of LTL properties as well as basic assertions and trace assertions (see \Cref{sec:props}).
These properties are useful for expressing logical requirements of a system that change over time---precisely the case for the \ln payment payment protocol.

\subsection{Scope and Threat Model}
In this work, we consider only the channel operation phase of the LN protocol, ignoring channel opening and closing.
This is because LN channels spend most of their time in channel operation, and this is the phase in which intermediate- to long-lived channels will send most of their messages.
A channel only needs to be opened and closed once, but once open, many payments can be made across it.
Our modeling of only part of the protocol is not an artifact of our methodology nor is it a reflection on computational restrictions for model checking the rest of the LN protocol.
It simply reflects the degree of manual work required to build a formal model, and is aligned with other works that model subsets of protocols~\cite{arai2014formal, chang2007formal,nguyen2014formal, phan2012analyzing, Modesti_Shahandashti_McCorry_Hao_2021, basin2018formal}.

We restrict our model to payments over a single channel.
We consider an adversary with full access to one of the peers on that channel,
including their cryptographic keys. The adversary can use the keys to encrypt, decrypt, sign or send  arbitrary messages, but cannot break the underlying cryptographic primitives.
Any attempt to forge a hash or signature will be detected.
While the LN uses encryption for all communication, we do not model this explicitly as we are not considering confidentiality or privacy properties. 
We consider this attacker model in lieu of the more conservative Dolev-Yao attacker~\cite{Dolev_Yao}, because the LN specifies that reliable communication channels must be used~\cite{bolt2}, and the cryptographic elements of the protocol render manipulation by a network adversary ineffective.

We do not consider partitions to the underlying blockchain network that could disrupt the consensus protocol, as that would leave the system vulnerable to much more powerful attacks.
However, we do allow that individual LN peers could become partitioned from the \ln and might be prevented from submitting transactions to the blockchain.

%% file: sections/model.tex
\section{Model Design}\label{sec:model}
To model the \ln, we built an event-driven finite state machine (FSM) of the LN protocol's payment flow stage.
We consider two types of events that initiate transitions: timeouts and receiving messages.
In response to an event, the FSM can either send a message, or update any state variables.
We present our state machine in \Cref{fig:fsm}.
We based our rigorous model on the published LN specification---the BOLTs~\cite{bolts}.
There are eleven BOLTs in the spec, but our work focuses on BOLTs two and three, which describe the channel operation behavior for peers that have already established a channel.
\addText{
Textual specifications lack rigor because natural language is ambiguous. 
Additionally, such specifications are often incomplete. 
When the textual specification in the BOLTs did not provide enough information or was ambiguous, we consulted the reference implementation (\lnd~\cite{lnd}) or the protocol developers directly. 
We also examined execution traces from our model and the Lighting Network implementation. 
As discussed in \Cref{sec:results}, the fact that the attacks we found with our model works on the real system implementation demonstrates the usefulness and fidelity of our model.
Further details about our model are available in \Cref{sec:modeldetails}.
}

\input{sections/fsm}

\subsection{Modeling Processes}
We model two LN peers that are directly connected via a single payment channel.
These two peers are modeled as independent processes within \spin.
Our model is written in {\spin}'s modeling language \promela.
We begin executing an LN peer process using the \promela \texttt{run}.
The two processes are ensured to start atomically by using {\promela}'s \texttt{atomic} keyword.
Atomicity at this stage ensures that, at least at the start, both processes are initialized and running.
Since we are modeling the payment flow, both peers must already have been running (to setup the channel) and likewise must not have closed the channel yet.
The starting state of both processes is \funded.
Aside from internal bookkeeping data, each peer tracks two state variables: the number of HTLCs opened by the local party and the number opened by the counterparty.

\subsection{Modeling Communication}
As our goal is to model payments between two peers, we must also model the network communication between those peers.
We model network communication via the \promela concept of \emph{channels} (not to be confused with payment channels).
In \promela, processes communicate by writing and reading from shared channels, which are queue-like data structures.
Writing to a channel enqueues a message, and reading from a channel dequeues a message.
Since we are using these channels to model network communication between distributed processes, we call a write to a channel a ``send'' event, and a read from a channel a ``receive'' event.
We model seven message types as detailed in \Cref{tab:msgs}.
These are all the possible messages that can be sent during payment except for a \texttt{WARNING} message, which we combine with the \texttt{ERROR} message as it can be used in all the same situations---the difference being that it carries slightly different information that might be of diagnostic use to the user.

Network communication in the LN is bidirectional.
We model bidirectional communication as a pair of channels.
Each peer sends to one channel and receives from the other with each peer only sending to the channel that the other peer receives from.
Concretely, peer $P_{A}$ sends to channel $C_{A}$ and receives from channel $C_{B}$, while peer $P_{B}$ sends to channel $C_{B}$ and receives from channel $C_{A}$.
In practice, a pair of LN peers will communicate over a single TCP connection, however, it is convenient for us to model communication as two unidirectional channels.
Splitting the channels up as such allows us to avoid modeling unrealistic network contention between the peers, such as one peer not being able to send a message because the other peer has already sent a message---two facts which are causally independent and would not cause issues over a TCP connection.

A \promela channel can be either buffered or unbuffered.
A process that sends to an unbuffered channel, a so-called \emph{rendezvous channel}, will block until another peer receives a message from the channel---logically, it is a queue of length zero.
Rendezvous channels are useful for modeling lock-step synchrony, because a process cannot continue its execution until its message is read.
However, the LN runs over the public internet where lock-step synchrony is not practical.

We opt to create buffered channels.
In \promela, a channel can include a fixed-size buffer, which indicates how many messages it can hold without causing the sender to block (i.e., a queue of length $n$, where $n>0$).
This models a bounded form of asynchrony, because the sender is able to continue execution before receiving confirmation that the peer has received the message.
To this end, we create channels with a buffer size of one.
This is sufficient to allow \spin to check a large number of otherwise unreachable states.
However, the state space complexity that the model checker explores grows exponentially with the channel size, so for practical constraints we keep the value to one. 
It is possible that there are states the protocol cannot reach unless there are more messages in flight, however, we believe the likelihood of this is small because the nature of this protocol is to try to maintain some degree of synchronization.
A party cannot get too far ahead of its counterparty, because it frequently needs information from the counterparty to proceed.

\subsection{Maximum Concurrent HTLCs}\label{sec:maxhtlcs}
A single channel cannot support more than \num{483} in-flight HTLCs.
According to BOLT \#8, the maximum transaction size in Lightning is \num{65}{kb}~\cite{bolt8}.
This maximum size is set to be less than the maximum \emph{standard} Bitcoin transaction size\footnote{This is more of a common practice than a hard rule. It is legal for larger transactions to be included in blocks, but many miners will not relay them.}.

Each commitment transaction sent between two peers must contain an output $\theta$ for each HTLC between those two peers.
If not, then there could be disagreement about the ratio of the payout from the funding transaction.
Given the effective maximum transaction size, the sizes of the required arguments of the transaction input $\tau.args$, and the script conditions $\phi$ for each output $\theta$, the maximum number of HTLCs concurrently in flight must be less than or equal to \num{483}.
In the protocol code and documentation, this is referred to as \maxhtlcs.

In our model, we set \maxhtlcs to be ten.
However, we do not believe that this constraint negatively impacts the fidelity of our results. 
In fact, a value of two for \maxhtlcs is sufficient to visit all system states for two peers, beyond that, new HTLCs only increment a counter.
While an increased counter is technically a new state, these new states do not add any conceptually interesting executions and thus are not worth the resulting state space explosion of using a \maxhtlcs of 483.
Our value of ten is enough to give confidence that no new interesting executions appear beyond a value of two \maxhtlcs, but still small enough to execute quickly.

%% file: sections/fsm.tex
\begin{figure*}[]
  \centering
  \includegraphics[width=\textwidth]{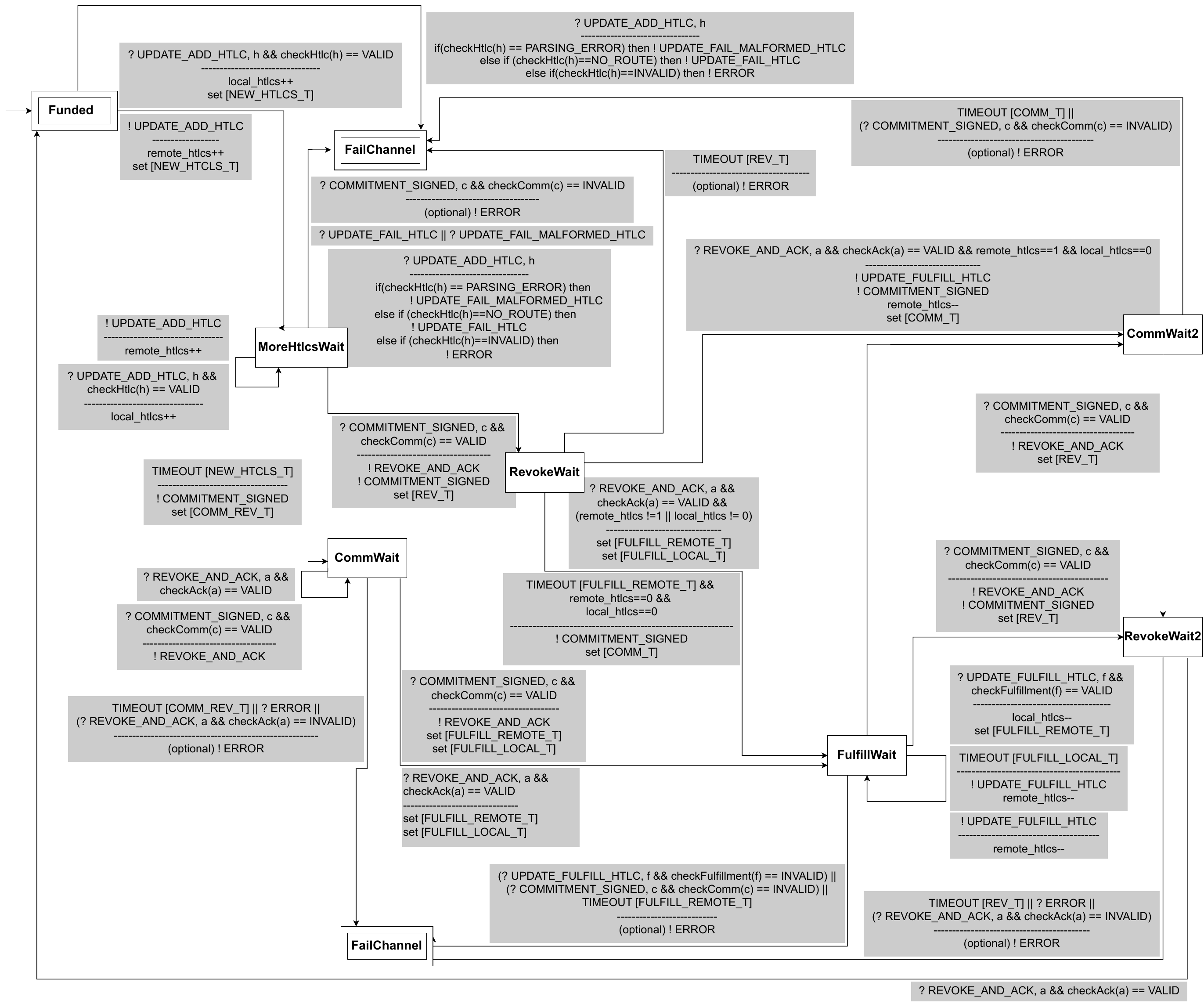}
  \caption{A finite state machine of the normal payment flow in the Lightning Network. Labels ending in ``\_T'' are timers. Execution starts in \funded. Both \fail states are equivalent and are duplicated for readability.}
  \label{fig:fsm}
\end{figure*}

%% file: sections/props.tex
\section{Properties}\label{sec:props}
We define the correctness of the \ln payment protocol in terms of certain properties it must satisfy.
Some of these properties, such as deadlock avoidance, are common among network protocols.
Others, however, are specific to the LN payment protocol.
These properties were either derived directly or were inferred from guarantees made in the BOLTs~\cite{bolts}.
In this section, we discuss these properties in depth.

\subsection{Detecting Deadlock and Livelock}
Model checkers can often verify if it is possible for a protocol to deadlock or livelock.
\emph{Deadlock} occurs when a run ends in a state in which no further steps can be taken but is not a legal end state.
\emph{Livelock}, on the other hand, is a type of non-progress cycle where execution can continue indefinitely without reaching a valid end state from which no further steps are possible.

To enable \spin to check for both deadlock and livelock, we must add \emph{meta-labels} to the \promela model.
Meta-labels are simply code-level labels that are prefixed with expected identifiers which confer additional meaning to those states.

The \spin model checker allows us to indicate which states are valid end states by prefixing a label with ``end'' in the \promela model.
Any execution that terminates at an end label is considered \emph{accepting}.
Our only labeled end states are \funded and \fail.
Likewise, \promela allows us to define progress states, which indicate that any infinite execution that includes the progress state infinitely many times is \buchi-accepting.
We denote progress states by prefixing a label with ``progress.''
Our only progress state is \funded.

\subsection{Claim Properties}
Beyond deadlock and livelock, we define five more properties for our model.
In the properties below, we define $P$ to be the set of peers.
We indicate that a peer is sending message $m$ to another peer by writing $! m$, and receiving message $m$ by writing $? m$.
If any message is legal, we write *.
If a set of messages $M$ is accepted except for a particular message $m$, we write $M/m$.
We have modeled some of the properties graphically as FSMs where applicable.

Our properties are either modeled as trace assertions which allow us to specify sequences of events that all accepting runs must always match, or \emph{linear temporal logic} (LTL), which allow us to specify sequences of states that all accepting runs must match.
We provide more details and background on types of properties and model checking in general in \Cref{apx:fm}.

\begin{figure}[h]
  \centering
  \begin{tikzpicture}
    \node[state, initial, circle] (A) {};
    \node[state, circle] (B) [right=3cm of A] {};

    \node[state, circle] (A) {}
      edge [->, bend left, above] node {! \revokeandackshort} (B);
    \node[state, circle] (A) {}
    edge [loop below, below] node[align=center] {
      ! */\revokeandackshort $\lor$ ? *
    } (A);
    \draw[->] (B)
    edge [bend left, above]
    node {? \commshort}
    (A);
  \end{tikzpicture}
  \caption{This automaton representing \Cref{prop:p1} should never generate an accepting run when executed synchronously with the model.}
  \label{fig:p1}
\end{figure}
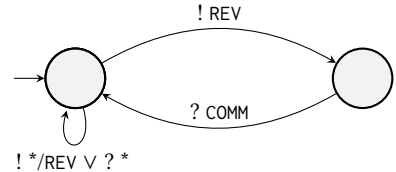

\begin{property}[Recourse Continuity]\label{prop:p1}
  A peer $p$ should never send a \revokeandack before receiving a \comm (\Cref{fig:p1}).
\end{property}

We model this property as a trace assertion, which we implement by claiming that no trace of events should exist where a peer sends a \revokeandack and then immediately receives a \comm.

A peer should always be able to unilaterally close a channel by submitting a commitment transaction to the blockchain.
However, if the local party revokes a commitment before receiving a new one, then they lose their ability to close without the possibility of penalty.
This is because commitments that the local party revokes can be contested by the counterparty providing the revocation key in $\tau.args$ to the spending script $\phi$.
The penalty allows the counterparty to spend all the UTXOs $\theta$ of the channel, accounting for the entire value of the channel.

Therefore, a peer should avoid putting itself in a position where the only commitment transactions it can submit to the blockchain can be contested (because they have all been revoked).
Note that a peer can only do damage to itself by prematurely sending a revocation.
A peer could also find itself in a similar position if it loses access to its private key.

\begin{figure}[h]
  \centering
  \begin{tikzpicture}
    \node[state, initial, circle] (A) {};
    \node[state, circle] (B) [right=2cm of A] {};
    \node[state, circle, accepting] (C) [right=2cm of B] {};

    \node[state, circle] (A) {}
      edge [->, above] node {? \commshort} (B);

    \draw[->] (B)
      edge [above] node[align=center] {
        ! \fulfillshort
      }
      (C);

    \draw[->] (A)
      edge [loop below, below] node {! * $\lor$ ? *}
      (A);
  \end{tikzpicture}
  \caption{The execution of an accepting run of the model should \emph{never} match this automaton representing \Cref{prop:p2a}.
  Concretely, a peer should never send a \fulfill directly after receiving a \comm---it should instead send a \revokeandack before the fulfillment.}
  \label{fig:p2a}
\end{figure}
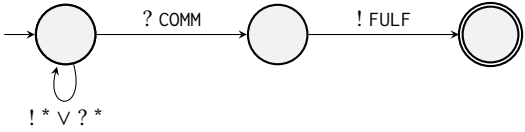

\begin{property}[Weak Channel Determinacy]\label{prop:p2a}
  A peer must not send or receive an \fulfill until it has sent a \revokeandack (\Cref{fig:p2a}).
\end{property}

A peer that does not revoke previous commitments retains the ability to close the channel with an old channel balance---a balance that does not include later payments.
This is a safety violation.
The counterparty would not be able to contest the commitment submitted to the blockchain because it does not have the revocation key that would have been included in the never-received revocation message.

An HTLC should not be fulfillable if there is more than one channel balance possible for a peer.
Such a situation would create ambiguity between the two channel peers about how much of the channel balance each of them controls.
We model this property as a trace assertion.

\begin{property}[Strict Channel Determinacy]\label{prop:p2b}
  If a peer receives a \comm, its next message must be a \revokeandack (\Cref{fig:p2b}).
\end{property}
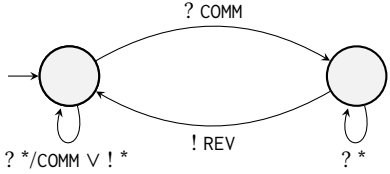
\begin{figure}[h]
  \centering
  \begin{tikzpicture}
    \node[state, initial, circle] (A) {};
    \node[state, circle] (B) [right=3cm of A] {};

    \node[state, circle] (A) {}
      edge [->, bend left, above] node {? \commshort}
      (B);
    \draw[<-] (A)
      edge [loop below, below] node[align=center] {
        ? */\commshort $\lor$
        ! *
      }
      (A);

    \draw[<-] (B)
      edge [loop below, below] node[align=center] {
        ? *
      }
      (B);

    \draw[->] (B)
      edge [bend left, below] node {! \revokeandackshort}
      (A);
  \end{tikzpicture}
  \caption{The execution of an accepting run of the model should \emph{always} match this automaton representing \Cref{prop:p2b}.}
  \label{fig:p2b}
\end{figure}

Like \Cref{prop:p2a}, failure to send a revocation can leave the channel in an ambiguous state where one peer has a full channel timeout period to decide which of the two channel states it wants to commit to.
A revocation should force the peer to choose a single state.

This property, again a trace assertion, is derived from the BOLTs explicit requirement that
``a receiving node [of a \comm message] MUST respond with a \revokeandack message.''
While \Cref{prop:p2a} precludes the possibility of a payment completing successfully without sending the requisite revocation, \Cref{prop:p2b} limits the possibilities further by saying that the immediate next message after a \comm must be a \revokeandack. 
Though the text does not directly state that a \revokeandack must be the \textit{immediate} next message, we infer this to be the intent in the BOLT as this particular line in the BOLT captures the requirements of a successful payment---where neither peer is at risk of losing funds.
The only other message besides \revokeandack that could make semantic sense towards completing a payment securely, would be another \comm, however, this just delays the problem because the counterparty (having already sent a commitment) cannot legally send another.

A violation of this property would mean that even though the payment was not successful, one peer could still spend the HTLC after an arbitrary delay (chosen during channel creation).

\begin{property}[HTLC Congestion]\label{prop:p4}
  A channel cannot have \maxhtlcs open. For a peer with $l$ locally opened HTLCs and $r$ counterparty-opened HTLCs, we can say using the LTL \emph{always} (written as $\square$).
  \begin{align*}
    \square (l + r < \maxhtlcs)
  \end{align*}
\end{property}

Going above the \maxhtlcs limit is unsupported behavior because it means that commitments and penalty transactions could become too large to be mined (\Cref{sec:maxhtlcs}); such an action should result in channel failure according to the BOLTs.
However, even having \maxhtlcs HTLCs open means that the channel cannot accept more HTLCs, thus blocking all the funds on the channel from use. 

\begin{property}[Payment Liveness]\label{prop:p5}
  An accepting run should always ($\square$) eventually ($\diamond$) terminate with the state $s$ in \funded or \fail, i.e., 
  \begin{align*}
    \square \diamond (s = \funded \lor s = \fail)
  \end{align*}
\end{property}

All payments should either complete successfully (end up in \funded) or follow a proper error path (end up in \fail).
Typically, an error leads to one or both of parties to force-close the channel, i.e., they close the channel without mutual agreement to do so. 
Repeatedly passing through \funded is allowed as this would constitute a \buchi acceptance cycle.

%% file: sections/results.tex
\section{Results}\label{sec:results}
\paragraph{Model Implementation and Execution} Our model is implemented in 300 lines of code. We model all seven message types that are used in channel operation (see \Cref{tab:msgs}), as well as all eight states shown in \Cref{fig:fsm}.

\paragraph{Environment}
All runs were executed on a 2021 Lenovo X1 Carbon ThinkPad with an 11th Gen Intel(R) Core(TM) i7-1165G7 processor clocked at \num{2.80}{GHz} and \num{32}{GB} RAM.
We use \spin version 6.5.2---the latest available at the time of our experiments.

\paragraph{Results Overview}
\addText{
\spin took approximately 0.1 seconds to verify each of the five properties and explored between 478 states (property 4) and \num{609502} states (property 5).
When checking for livelock and deadlock, \spin explored \num{288139} states, and the highest memory usage in any of the runs peaked at 175 MB.
}
Our verification with \spin did not report any deadlocks or livelocks.
However, of the five properties we defined (\Cref{sec:props}), we found that two of them can be violated in accepting runs.
Each of these violations can result in an attack, one of the attacks is known in the literature, the other is novel to the best of our knowledge.
We discuss the property violations and attack mechanisms in detail below.

\subsection{Uncovering Known Attacks}
\paragraph{A1: Congestion Attack}
Our system correctly captured Mizrahi and Zohar's congestion attack~\cite{mizrahi2021congestion}.
This is the direct implication of the \ln protocol violating \Cref{prop:p4}, which states that a channel should never have \maxhtlcs HTLCs open.

In this attack, a malicious peer sends new {\add}s until hitting the \maxhtlcs HTLC limit.
This action is similar to a \emph{griefing} attack. 
A griefing attack occurs when an honest user is unable to route a payment over a channel because that channel cannot accept additional HTLCs.
This is usually because the full amount of funds on the channel are already allocated to prior in-flight payments.
However, in this case, the cause is that the number of HTLCs has hit the mandated hard limit.
Griefing has been studied extensively in previous works~\cite{perez2020lockdown, mizrahi2021congestion, sprites, Egger_Moreno-Sanchez_Maffei_2019,Aumayr_Abbaszadeh_Maffei_2022}.

What makes this property (\Cref{prop:p4}) interesting is that peers are explicitly allowed to open \maxhtlcs HTLCs~\cite{bolts}.
Whatever the limit is set to, the protocol will naturally permit that many HTLCs, so, in fact, it is expected that the model violates this property.
Nonetheless, this violation serves as additional confirmation that our model is accurate.
For a discussion on mitigating this attack, we refer the reader to \citet{mizrahi2021congestion}.

\subsection{Uncovering Novel Attacks}
Our verification of the \ln payment protocol led us to what we believe is a novel attack against the payment protocol.
\spin did not uncover the attack in its entirety, but the full attack is only a slight extension to the property-violating trace.
We describe the attack mechanism and implications below.

\begin{figure}[h]
     \centering
     \includegraphics[width=\linewidth]{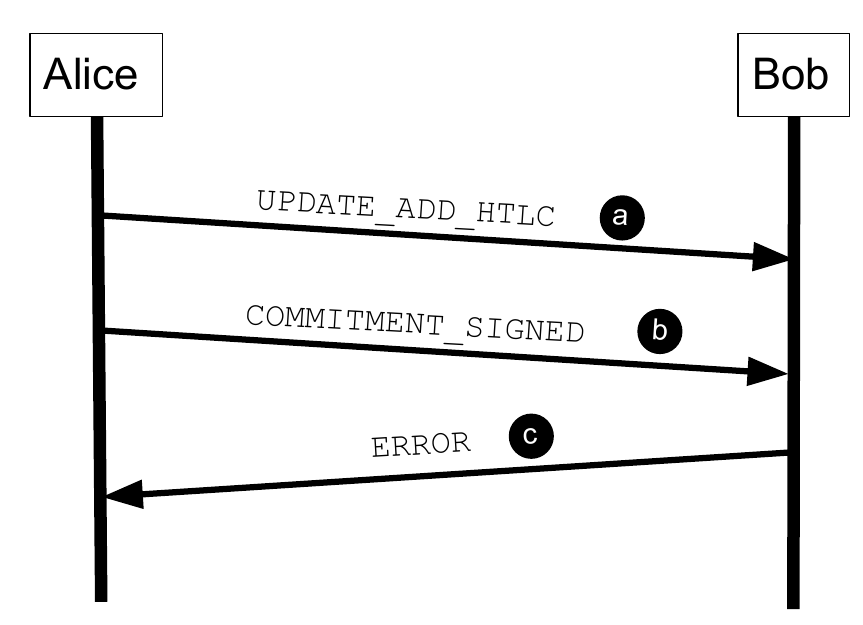}
     \caption{The shortest trace that violates \Cref{prop:p2b}.}
     \label{fig:p2bviolation}
\end{figure}

\paragraph{Property Violation}
There are, in fact, an infinite number of accepting runs that violate \Cref{prop:p2b}---a result of the payment protocol being an infinite \buchi automaton.
However, the shortest violating trace is straightforward as is shown in \Cref{fig:p2bviolation} in a payment between two LN peers, Alice and Bob.
While it is impossible to enumerate all violating traces, we suspect that all such violations share a common cause.

The shortest trace in which this violation occurs happens after one peer, Alice, adds a new HTLC and then immediately sends a commitment.
The channel counterparty's next message must be a \revokeandack according to the BOLTs, however, the BOLTs also explicitly state that

\begin{quote}
``From the point of waiting for \texttt{CHANNEL\_READY} [a message sent during channel establishment] onward, either node MAY send an \error and fail the channel if it does not receive a required response from the other node after a reasonable timeout.''
\end{quote}

This is, in fact, an ambiguity in the specification, because a peer cannot \emph{always} send a \revokeandack if it also always allowed to send an \error and fail the channel.

As discussed in \Cref{sec:bkg:chan_op}, during a \ln payment, each peer must successively commit to new states---this is the core process in sending payments off-chain.
In order to make sure that \emph{at all times} each peer is able to claim its channel funds on the blockchain, each peer must always maintain at least one valid \comm message.
The implication of this is that at some points during protocol execution, one or both peers may have \emph{two or more} legal commitments that they could redeem on the blockchain.

Having two legal channel states can lead to ambiguity in how much the channel should payout to each party, if the payment protocol does not reach its prescribed conclusion.
In most cases and in successful payments, this is not a problem, because after receiving a new commitment, a peer revokes its previous commitment with a \revokeandack.
The revocation thus limits the peer again to only a single commitment transaction that it can safely (i.e., without fear of penalty) submit to the blockchain.

\paragraph{A2: \attack Attack}
The violation of \Cref{prop:p2b} showed that the protocol can terminate in an undesirable state.
This observation led us to consider what would happen if there were a network partition after Alice sent message \step{b} in \Cref{fig:p2bviolation}. 
In general, a peer cannot be certain that its counterparty received the commitment until it receives an acknowledgment in the form of a \revokeandack message.
The implications of this are critical, which becomes clear when we consider how this could play out in a point-of-sale situation.

Consider the scenario where Alice is at a brick-and-mortar store trying to purchase product $\mathcal{P}$ via transaction $\mathcal{TX}$ from Bob.
Prior to the payment, Alice's most up-to-date commitment $C2_{B}$ and Bob's most up-to-date commitment $C2_{A}$ agree on the payout amounts available to both Alice and Bob---these are encoded in the commitments' respective UTXOs.
Both $C2_{B}$ and $C2_{A}$ can accept the same UTXO input $\tau$, if they are force-closing a channel.
In this case, $\tau$ is equal to the output of the channel funding transaction ($tx_{fund}.\theta$).
Since, both of these transactions spend the same UTXO, they cannot both be included in a block.
This is normally not a problem as both parties are storing these commitments locally and only publish them when they want to close the channel.
However, the ambiguity can arise if Alice then sends a new HTLC to Bob (message \step{a}) followed by a new commitment $C1'_{A}$ (message \step{b}), and the network partitions sometime after Alice sends message \step{b}.
From the BOLTs, we can deduce several ways to resolve this ambiguity.
Most of these resolutions result in both parties either agreeing that the transaction completed successfully or agreeing that it did not.
Below however, we highlight two resolutions that lead to disagreement on the state of the channel.

\paragraph{Outcome 1: Honest Bob, Malicious Alice}
If Bob \emph{did} receive the new commitment $C1'_{A}$ and he is honest, he may be inclined to give product $\mathcal{P}$ to Alice and consider the payment complete.
After he hands over product $\mathcal{P}$, Alice can force close the channel with the latest commitment $C2_{B}$ she has access to.
That commitment $C2_{B}$ spends the funding output $tx_{fund}.\theta$, thus precluding Bob from publishing his updated commitment $C1'_{A}$ in which he receives payment from Alice for $\mathcal{P}$.
\addText{
In this case, Bob, the seller, has received a commitment but Alice, the buyer, has not. 
The seller may then view the payment as valid and hand over the product before the buyer is committed to completing the payment.
This is allowed because the buyer has not yet revoked the previous channel balance.
This is analogous to buying something with a credit card and then reporting it as fraudulent spending to the creditor---except in this case, there are no laws or protections for the buyer or seller.
}

\paragraph{Outcome 2: Honest Alice, Malicious Bob}
Consider an alternative scenario where Bob lies and claims that he \emph{did not} receive the commitment, which again, Alice cannot confirm or deny.
Due to the partition, both Alice and Bob agree out-of-band (perhaps verbally) that the payment is a failure and Bob does not give Alice product $\mathcal{P}$.

Subsequently, Bob force closes the channel with the commitment $C1'_{A}$ that he did, in fact, receive.
Bob can immediately spend the HTLC output because he already has the payment preimage,
which he can provide as one of the $\tau.args$ to the HTLC output script $\theta.\phi$.
\addText{
In this case, Alice, the buyer, has received a commitment but Bob, the seller has not, the buyer then might view the payment as invalid, though they have already committed to making the payment.
This ambiguity is not discussed in the BOLTs and can leave the buyer vulnerable to lost funds.
}
In practice, there are delays to consider before the outputs can be spent into Bob's wallet.
We will discuss those details more in the following section.

\paragraph{Impact and Practicality}
If the network partitions after Alice sends message \step{b}, then there are no guarantees that the resolution will be safe for both Alice and Bob.
That said, if both Alice and Bob are honest, then a safe resolution is indeed possible.

Eventually, either the HTLC will timeout, or the force closing transaction will be mined at which point the outcome will be determined and known to both Alice and Bob.
However, waiting for such an action or timeout may be unreasonable.
HTLC timeouts (called \texttt{cltv\_expiry}) can be negotiated during channel setup.
Most users, however, simply rely on the defaults used in their implementation~\cite{Zabka_Foerster_Schmid_Decker_2022}.
The most popular implementation of an LN peer, capturing about 87\% of the market~\cite{Zabka_Foerster_Schmid_Decker_2022}, is \lnd~\cite{lnd}.
The default \texttt{cltv\_expiry} timeout of \lnd is forty blocks (approximate clock time: six hours).
If Alice is buying, say, a slice of pizza, she may prefer not to wait in the pizzeria for six hours until she is certain she has not been duped.

In practice, there are certain real-world expectations that may make such an attack infeasible.
For instance, in Outcome 1, if Bob does redeem the HTLC without giving Alice product $\mathcal{P}$, Alice could simply return to the store and request renumeration, or post a review to alert others of unsavory business practices.
In most cases, a business will have an economic incentive to behave honestly.

In the second attack outcome, Alice's safest action is to force close the channel herself, which she must do before Bob.
This is a race condition whose outcome is subject to the whims of miners. In the worst case, it opens the door for bribery and frontrunning~\cite{flashinthepan,flashboys2,hehtlc,madhtlc}.
For instance, an adversary could bribe a miner to mine their force close transaction, which would yield economic benefit to both as long as the bribe amount is less the HTLC payout.
Nonetheless, if this network partition is not a random occurrence but rather a concerted effort by Bob or someone else, it is not a stretch to imagine that Alice may not have the network connectivity required to publish the force close transaction at all.

Ultimately, these risks will be present anytime there is an exchange of physical goods, however, the ambiguity brought about in Outcomes 1 and 2 make these attacks a little bit easier, thus lowering the bar to such behavior.

\subsection{\attack Proof-of-Concept}\label{sec:poc}
We have validated the above \attack Attack on the popular \lnd implementation.
While we show that it is exploitable in an operational implementation, the following proof-of-concept was performed in a local, firewalled environment---no transactions or channels were observable or accessible by anyone outside of the authors' environment, and no real bitcoins changed hands.

The core intuition of this attack is that when a network fault occurs, and there was an information asymmetry beforehand between the two channel peers, the protocol can lead to undefined behavior. 
The specific asymmetry is that one peer has two valid commitments, while the other only has one.
To show that this attack is, indeed, exploitable, we show that the balances dictated by all three commitments ($C2_{A}$, $C2_{B}$, and $C1'_{A}$) are spendable if the respective commitment were published on the blockchain.

\paragraph{Setup}
Our environment consists of four components: two LN nodes, one Bitcoin node, and one orchestration client.
The LN nodes both used a modified \lnd implementation.
The modification allowed us to simulate a network partition after sending the first commitment (message \step{b} in \Cref{fig:p2bviolation}).

The Bitcoin node we used was running \texttt{btcd}, a Go implementation of the Bitcoin protocol.
The Bitcoin node formed its own Bitcoin network with a hash difficulty of one.
This means that finding the nonce that satisfied the proof-of-work challenge was trivial: any hash greater than one sufficed.
This was important because, as will be clear below, certain outputs required a block to come to maturity 
before they could be spent---thus we had to sometimes mine upward of \num{1000} blocks.
With such a low proof-of-work difficulty, this was a quick process.

The orchestration client was only used to interact with the other three nodes using their RPC interfaces.
It served as a command center.
At the conclusion of each of the following test cases, we deleted our setup and re-initialized it for the next scenario, ensuring that the experiments were executed independently. 

\paragraph{Attack Execution}
The concurrent commitments ($C2_{A}$, $C2_{B}$, and $C1'_{A}$) all spend the same UTXO, so only one can ever be published on the blockchain.

Confirming the validity of the first two commitments, $C2_{A}$ and $C2_{B}$, is trivial.
In each of two separate test runs, we executed the \lnd command to force close the channel from each of the peers, respectively.
In both cases, this generated a channel closure transaction and transmitted it to our running Bitcoin node.
The LN protocol has a built-in delay before the force closure transaction becomes spendable. We mined \num{1081} blocks to bring the transaction to maturity, so that it could, itself, be included in a block and mined.
\addText{The \num{1081} blocks are variable. For channels with fewer funds, it can be as low as 144 blocks\footnote{\url{https://github.com/lightningnetwork/lnd/blob/07b6af41dbe2a5a1c85e5c46cc41019b64640d90/funding/manager.go\#L78}}. This is decided by the peer who sends the \ttt{OPEN\_CHANNEL} message that initiates the channel establishment phase. While this is the time that a peer needs to wait to receive the funds in their wallet, once a peer has force-closed a channel, the payout is a forgone conclusion. As our attacker has not yet shared their revocation key, the penalty payout is not an option.}
Upon mining a block with this transaction, the a new UTXO was created which could only be spent to the force-close-initiator's wallet.

We also showed that spending the third commitment ($C1'_{A}$) is possible.
This commitment is structurally different than the other two ($C2_{A}$ and $C2_{B}$).
Those commitments have a straightforward spending path, because to receive them, both parties have to prove that they have received the first type of commitment ($C1_{A}$ and $C1_{B}$, respectively).
They were able to prove this by sending a \revokeandack.
Commitment $C1'_{A}$ (sent by Alice, held by Bob) has more built-in spending delays to protect Alice and allow her to redeem funds with the preimage, if she acquires it.
Whereas Bob, who already has the preimage, must wait until the \texttt{ctlv\_expiry} expires.
The delays also protect Alice by giving her time to submit the revocation key if the commitment had already been revoked.

\begin{figure}[h]
     \centering
     \includegraphics[width=\linewidth]{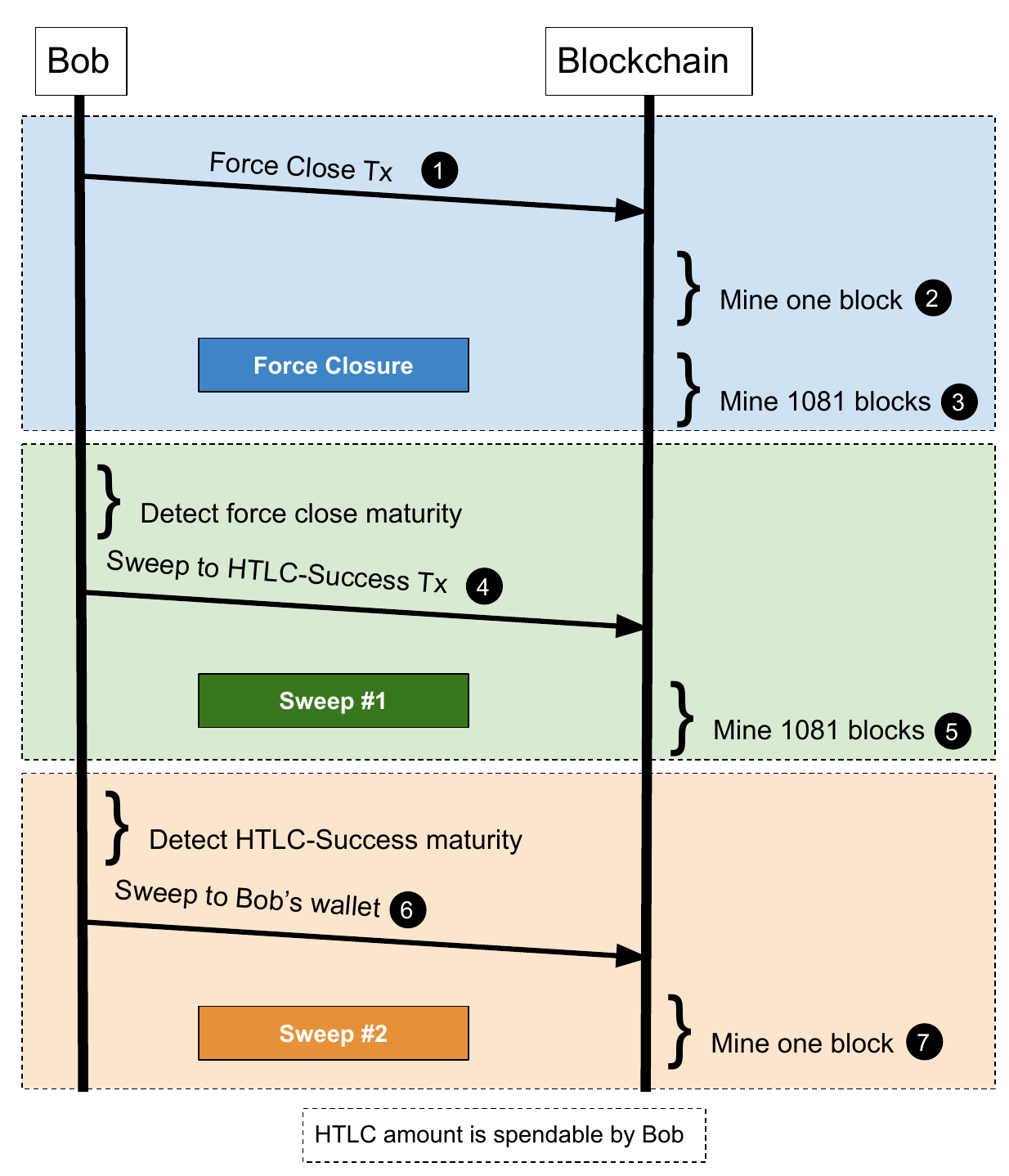}
     \caption{Steps required for Bob to retrieve funds in the \attack attack.}
     \label{fig:cashout}
\end{figure}

We showed that Bob can retrieve the funds using the following steps (\Cref{fig:cashout}):
\begin{inparaenum}[(1)]
  \item [\step{1}] submit the force close transaction to the blockchain;
  \item [\step{2}] wait for one block to be mined to get the force closure included in the blockchain;
  \item [\step{3}] wait for an additional \num{1081} blocks to be mined to bring the channel closure to maturity;
  \item [\step{4}] wait for one minute allowing Bob's \lnd node to detect that the channel closure is ``mature'', i.e., spendable, at which point his node submits a transaction to the blockchain network that includes the preimage and \emph{sweeps} the channel closure funds into a new temporary contract (called an \emph{HTLC-success transaction});
  \item [\step{5}] wait for \num{1081} blocks to be mined to bring the HTLC-success transaction to maturity;
  \item [\step{6}] wait for one minute allowing Bob's \lnd node to detect that the prior sweep transaction is mature, and submit the second round sweep transaction to the network;
  \item [\step{7}] wait for one block to be mined,  which confirms the sweep and sends the HTLC funds to Bob's wallet. 
\end{inparaenum}

\subsection{Mitigation}
Forming a payment channel with a peer is an economic commitment that a peer should not take lightly.
Prior works have suggested that a payment channel network's reliability may be improved by incentivizing alternative connection patterns~\cite{Weintraub_Nita-Rotaru_Roos_2021}---for example encouraging peers to form channels with only parties they trust, rather than whoever can offer the shortest routing paths.
A trusted peer is less likely to try to cheat its counterparty.

Watchtowers are a critical tool in the LN ecosystem~\cite{pisa, dryja2016unlinkable}.
Their job is to watch the blockchain at all times and alert users if a peer attempts to publish an already revoked commitment.
While useful in the case of revocations, watchtowers will not help against this attack, because by the time a force close transaction is published, it is already too late.
Additionally, the watchtower will be unaware of how any real-life exchange of goods may have turned out.

Some LN clients have included features that shorten the window of vulnerability.
These features, however, are independent of the BOLTs.
For instance, \lnd tracks local party, counterparty, and pending commitments independently\footnote{\url{https://github.com/lightningnetwork/lnd/blob/bbbf7d33fb1527acebb44e2a69d16fbcf24cc2fa/contractcourt/channel\_arbitrator.go\#L1779-L1880}} so that it knows which payments have not fully completed the protocol.
The latest versions of \lnd also include a signal to higher-level applications when they can safely consider a payment complete. 
Only upon such confirmation should the operator of a peer arrange a transfer of goods or services, thus reducing ambiguity for failed payments.

The \lnd implementation has also addressed the case when a counterparty HTLC is close to the \texttt{final\_cltv\_delta} timeout (the absolute timeout relative to the block height at the time of the invoice).
In this case, the sender's \lnd client automatically spends the HTLC back to its own wallet.
This naturally avoids the race-condition that occurs after the \texttt{final\_cltv\_delta} timeout expires.

Another defense in \lnd protects the HTLC sender if the channel counterparty attempts to spend the HTLC using the payment preimage.
To protect the HTLC sender, \lnd monitors the public mempool and can detect if a preimage-supplied closing transaction has been submitted.
In that case, it knows the counterparty is trying to cheat and can force close the channel preemptively.
This is possible because timeouts are asymmetric---the sender can close an HTLC with a preimage immediately, whereas the receiver must wait until a timeout has elapsed.

These defenses, while useful, are not discussed in the BOLTs and thus less actively developed clients may still be at risk.
Additionally, they do not help a client who is fully partitioned from the network, such as shown in our PoC (\Cref{sec:poc}).

%% file: sections/related.tex
\section{Related Work}
Protocols, on the blockchain and otherwise, have been evaluated using a variety of techniques in the literature.
Popular among these techniques are fuzzing, theoretical analysis, automated theorem proving, and model checking.

\paragraph{Fuzzing} Fuzzing uses heuristics to generate protocol events that are likely to cause problems.
It has previously been used to analyze network  protocols~\cite{snake,resolfuzz,tcpwn}, and in the blockchain space it has been used for analyzing smart contracts~\cite{Liu_reguard_2018, Jiang_contractfuzzer_2018, confuzzius}.

Fuzzing is useful for finding problems in protocols with complex state spaces, but it has three main drawbacks for checking protocols:
\begin{inparaenum}[(i)]
  \item it does not offer any guarantees that it will find any problems if they exist,
  \item it is restricted to finding faults with obvious external effects (e.g., fail-stop faults), and
  \item it requires an implementation to test against.
\end{inparaenum}
Fuzzing is generally not focused on ensuring semantic correctness of a protocol, only on making sure that implementations of the protocol do not crash.
The inability for fuzzing to guarantee correctness limits its usefulness for analyzing a financial application such as the \ln---especially when stronger guarantees are possible with other approaches.

\paragraph{Theoretical Analysis}
The security of the \ln has been analyzed theoretically from the perspective of the underlying ledger functionality and cryptographic primitives~\cite{Kiayias_Litos_2020} as well as from a game-theoretic perspective~\cite{Rain_game_2023, brugger2022automating}.
The security of the Lightning network's cryptographic protocols was proven in the universal composability setting~\cite{canetti2007universally}, which has also been used to analyze alternative payment channel network designs~\cite{malavolta2017concurrency,dziembowski2018general} with slightly different functionality than Lightning.  
%\stef{added the previous sentence on crypto}
In terms of game theory, \citet{Rain_game_2023} provide behavioral recommendations for rational parties to maximize their utility during channel closures and also verify the wormhole attack (\citet{wormhole}) in multi-hop payments.
However, it is difficult to prove correctness of a complex protocol with only theory, because a high-fidelity model of a protocol often contains too many possibilities for a human to reason about.

\paragraph{Automated Theorem Proving} Formal methods techniques offer the possibility of exploring the exhaustive set of legal protocol executions.
Correctness claims may be defined for the protocol in question, and automated theorem proving tools guarantee finding any instances of the claims being violated, if possible.

Theorem provers have been used in prior work for analyzing smart contracts~\cite{Amani_verifying_2018, Annekov_concert_2020, Sun_formal_2020, grishchenko_ethertrust_2018} as well as in a variety of network protocols like TCP~\cite{Bishop_Fairbairn_Mehnert_Norrish_Ridge_Sewell_Smith_Wansbrough_2019}, TLS~\cite{bhargavan2017verified, delignat2017implementing}, and Bluetooth~\cite{arai2014formal, chang2007formal, nguyen2014formal, phan2012analyzing, wu2022formal}.
They have also been applied to gossip protocols~\cite{gossipsub_sp2024} and cryptographic protocols such as Signal~\cite{kobeissi2017automated} and Noise~\cite{girol2020spectral, kobeissi2019noise}.

To the best of our knowledge, they have not been used for any payment channel network protocols like the \ln.
Additionally, while theorem proving is useful for confirming the correctness of algorithms, it is less useful for exploring behaviors of transition systems such as protocols specified as finite state machines.
Theorem provers also require an executable model, which may not be readily available.

\paragraph{Model Checking}
Model checking has been used to successfully find faults in many protocols~\cite{von2020automated, maggi2002using, lowe1995attack, lowe1996breaking, basin2018formal, cremers2020clone, cremers2017comprehensive, cremers2019prime}.
It has also been used extensively in the context of blockchain smart contracts. \citet{Bai_formal_2018} and \citet{Osterland_model_2020} used the \spin model checker to formally verify security properties in \emph{linear temporal logic} (LTL) for smart contracts.
Other works~\cite{Almakhour_formal_2023, Nehai_model_2018, Mavridou_verisolid_2019} apply model-checking to smart contracts by verifying security properties defined using \emph{computational tree logic} (CTL).
\citet{Nam_formal_2022} additionally model and verify game-theoretic properties by treating smart contracts as a two-player game and using \emph{alternate-time temporal logic} (ATL) to define security properties.

\citet{Modesti_Shahandashti_McCorry_Hao_2021} used the OFMC symbolic model checker to verify Bitcoin's BIP70 Payment Protocol~\cite{bip70}---which, while valuable, is qualitatively different than the \ln's payment protocol.
More similar to our work, \citet{Grundmann_2021} model checked the \ln protocol for single-hop transactions (i.e., payments that do not traverse multiple channels) including the underlying blockchain mechanisms.
\addText{In an extension of this work, \citet{Grundmann_Hartenstein_2023}} model multi-hop payments but still do not find any property violations.
The authors of these two works used TLA$^+$ model checker.
While they did model all phases of the LN protocol, they are missing significant details---likely the reason they report finding no property violations.
\addText{
Their works differs from ours in three main ways.
First, they only model a single payment per channel, thus not considering effects of concurrency. 
Second, they do not model error conditions, which was necessary for our discovery of both vulnerabilities and may be why they did not find any.
Finally, their single-hop property models only the correctness of the revocation mechanism, while their multi-hop property tracks only channel balances and does not capture even well-known attacks (e.g., wormhole). 
}

%% file: sections/conclusion.tex
\section{Concluding Discussion}
The \attack attack is made possible because of a transient information asymmetry between the two participating peers.
And although this design results in a weakness that our attack exploits, it is in fact intentional.
It reflects an effort by the designers to permit payments to occur asynchronously---even while one party is temporarily offline.
As a result of this goal, there is no possible complete defense against the attack for this protocol as it presently relies on successive two-phase commits.
This is because the information asymmetry will always exist if the commitment exchanges do not occur concurrently---something that can never be guaranteed.

A protocol that avoids two-phase commits altogether, such as in \citet{blitz}, could effectively avoid this attack vector. 
However, such schemes introduce additional complexities and may increase block size due to the need to include full rollback UTXOs in every transaction.
Transaction size is a sensitive issue in the blockchain space and might make this solution unpopular~\cite{Rizzo_2016}.
Synchronized clocks are also not a satisfactory solution as such an infrastructure is expensive and thus not suitable for peer-to-peer applications.
Further, in the best case they only narrow the time window in which this protocol is vulnerable.

One last mitigation worth considering would be to make the force closure transaction not minable until after a delay (i.e., until a certain block height). 
By having it be immediately minable, Alice could lock in her refund by getting the original commitment mined.
However, this would make channel closure---an already slow process---even slower.

While it is surely a net positive that \lnd has implemented many mitigations for the \attack attack, the \ln is meant to be an open network with safety guarantees for any participant following the protocol specification in the BOLTs.
Given the ease with which an adversary can fingerprint an implementation type~\cite{Zabka_Foerster_Schmid_Decker_2022}, an attacker could target vulnerable implementations.
Thus it is imperative that the BOLTs are explicit about when race conditions might occur, and suggest possible mitigations.

%% file: sections/acks.tex
\section*{Acknowledgements}
Ben Weintraub would like to thank Lisa Oakley and Max von Hippel for many insightful discussions about model checking.
Parts of this work were sponsored by NSF grant SaTC 1801546.
Disclaimer: Any opinions, findings, and conclusions or recommendations expressed in this material are those of the author(s) and do not necessarily reflect the views of the National Science Foundation.

%% file: sections/formal-methods.tex
\section{Formal Methods}\label{apx:fm}

Network protocols are complex systems with behaviors that are often difficult to reason about.
They are defined by the semantics of the messages sent between peers in the network, and how those peers respond to various events.
Events include: receipt of a message, completion of an internal computation, or a timeout elapsing.

Most open network protocols publish some form of specification document (e.g. an RFC~\cite{tcprfc, dnsrfc, quicrfc}) that dictates what actions peers can and should take in response to possible events.
However, it is difficult to know if the protocol definition is itself flawed.
The protocol definition may, in fact, cause all correct implementations to end up in situations where core goals of the protocol are violated.

Formal methods  allow engineers and researchers to explore more protocol execution paths than they would see during normal operation and traditional testing.

\subsection{Model Checking}\label{sec:modelcheck}
A formal methods tool of particular interest to us is \emph{model checking}.
A model checker can search through all possible executions of a protocol and is guaranteed to report any violations of modeler-defined properties.
This global set of executions is the \emph{state space}, where we define a \emph{state} to mean a unique set of values of all protocol variables~\cite{Baier_Katoen_2008}.
The model checker either reports that the properties are satisfied in all executions, or it reports the exact trace of events that cause the property violation.
The downside of model checking is that it can be subject to \emph{state space explosion} in which there are too many possible executions of a protocol to check them all.
However, modern model checkers are designed to massively reduce the state space by quickly ruling out impossible and redundant states, thus making them practical for many use cases~\cite{Baier_Katoen_2008}.

A model checker accepts two inputs:
\begin{inparaenum}[(i)]
  \item a protocol modeled as a finite state machine (FSM), and
  \item a set of properties that the protocol must satisfy.
\end{inparaenum}
It returns any property violations of the protocol including the exact trace of events that led to the violation.

\paragraph{Protocol Model (Input-I)}
The model checker accepts as input a \emph{finite state machine}, $M$, which consists of the tuple $(S,s_0,L,T,F)$, where $S$ is a set of states, $s_0$ is the starting state ($s_0 \in S$), $L$ is a set of labels, $T$ is a set of transitions ($T \subseteq S\times L \times S$), and $F$ is a set of final states ($F \subseteq S$).

\paragraph{Protocol Properties (Input-II)}
Exploring a model has only limited value without properties to check the model against.
We define properties as invariants that must remain true in all accepting runs of the protocol.
A run is an ordered, potentially infinite, set of transitions where the starting state is $s_0$ and the second state in any given transition matches the start state of the next transition, e.g., $\{(s_0, l_0, s_1), (s_1, l_1, s_2), (s_2, l_2, s_3)...\}$.
We describe a run as \emph{accepting} if the run is both finite and the last transition of the run, $((s_{n-1}, l_{n-1}, s_n))$, satisfies the property $s_n \in F$.
We consider a run \emph{\buchi-accepting}~\cite{buchi} if it is an infinite run that includes the state $s_f \in F$ infinitely often.

Some properties are implied, and will be checked by model checkers automatically, for example, ensuring that all runs are accepting, i.e., there are no deadlocks.
Some model checkers also allow modelers to define their own properties, which may include application-specific notions of safety, liveness, and fairness.
Model checkers can verify properties encoded in any of several forms of logic.
We focus on four forms in particular: basic assertions, never claims, trace assertions, and linear temporal logic (LTL).

\begin{itemize}
  \item[\textit{Basic assertion}.] Statements interwoven into the model execution that check if some condition is true. An assertion makes some claim about the value of system variables.
  \item[\textit{Never claim}.] Whereas assertions report specific states that should never exist during program execution, never claims assert fragments of executions which should never be included in an accepting run.
  A never claim is, itself, a fully defined FSM.
  \item[\textit{Trace assertion}.] Allow the modeler to specify sequences of events that all accepting runs must always match.
  \item[\textit{LTL}.] Allows the modeler to specify sequences of states that all accepting runs must match.
\end{itemize}

\paragraph{Property Violations (Output)}
An execution of a model checker against the provided properties can yield two possible outcomes.
On one hand, it might confirm that the properties hold in all accepting runs.
On the other hand, it might report that it is possible for the property to be violated.
In this case, the model checker will return the trace of events that led to this violation.

\paragraph{Semantics and Execution of Model Checkers}
A model checker's exploration is driven by the verification of properties.
Basic assertion properties operate similarly to code instrumentation.
That is, they are checked at exactly the point in the model where the assertion is specified, and will report a violation if the assertion statement is not satisfied.
Never claims, on the other hand, are verified by searching through both the model FSM and the claim FSM in parallel, i.e., each event causes independent transitions in both FSMs.

If any event causes the model FSM to enter a state not defined in the claim FSM, then a violation is reported.
As LTL and trace assertions are transpiled to never claims at runtime, this semantics applies to them as well.

%% file: sections/model-details.tex
\section{Model Details}\label{sec:modeldetails}
\addText{
Our model consists of eight states and nineteen transitions.
Note that \Cref{fig:fsm} has two boxes labeled \fail, this is for clearer visualization, but both boxes represent the same state.
All executions start in the \funded state, and all accepting runs must terminate in either the \funded or \fail states.
}

\addText{
All states have a transition to \fail, as specified in the BOLTs. 
These transitions represent executions where the previously received message was invalid or a timeout elapsed.
If the received message is an \add, then there are two additional error messages possible depending on the type of error: \failhtlc or \malformedhtlc.
Timer variables are indicated with the suffix ``\_T''.
There are timers for all ``Wait'' states in which a peer is expecting a particular message to be able to complete the protocol successfully. 
Timers begin when a message is received leading to a ``Wait'' state.
A peer transitioning to \fail may optionally send an \error message to the other peer.
}

\addText{
Aside from timers, we track two additional state variables.
They are \texttt{local\_htlcs} and \texttt{remote\_htlcs}.
They represent the number of HTLCs the local peer and the remote peer presently have open, respectively.
}

\addText{
We define four functions for succinctly representing message parsing: \ttt{checkHtlc()}, \ttt{checkComm()}, \ttt{checkAck()}, and \ttt{checkFulfillment()}.
These each accept the most recently received message and return either \texttt{VALID} or \texttt{INVALID}.
}

\addText{
In general, the payment flow executes as follows.
First, one peer sends an HTLC message to the other peer in the form of an \add message.
This may be followed by more HTLCs from either peer.
Next, one peer will send a commitment (\comm) message, and the other peer will respond with an ack and revocation secret (\revokeandack).
This is followed by the other peer sending their commitment, and the counterparty sending their revocation.
Upon completion of the commitment/revocation exchange, one peer will send a \fulfill message indicating that the payment is valid and agreed upon.
Then the two peers will exchange another commitment/revocation pair.
This second commitment/revocation exchange is used to shorten the commitment message to take up less block space---this becomes possible at this point because less contract logic is needed to protect the peers from malicious behavior after both peers have sent their revocations.
Upon receiving the second round revocation, the payment is considered complete and returns to the \funded state.
}